\PassOptionsToPackage{table}{xcolor}
\documentclass[nojss]{jss}

\usepackage{orcidlink,thumbpdf,lmodern}
\usepackage{amsmath,amssymb}
\usepackage{float}
\usepackage{subcaption}
\usepackage{booktabs}
\usepackage{tikz}
\usetikzlibrary{arrows.meta,positioning,fit,backgrounds}

\author{Robert W.~Krause~\orcidlink{0000-0003-4288-4732}\\National Institute of Eduaction, Nanjang Technical University, Singapore}
\Plainauthor{Robert W. Krause}

\title{\pkg{int3ract}: Johnson--Neyman Technique and Its Three-Way
       Extension for Frequentist and Bayesian Models in \proglang{R}}
\Plaintitle{int3ract: Johnson-Neyman Technique and Its Three-Way Extension
            for Frequentist and Bayesian Models in R}
\Shorttitle{\pkg{int3ract}: Johnson--Neyman Technique in \proglang{R}}

\Abstract{
  Interaction effects are ubiquitous in applied statistical modeling, yet
  their meaningful interpretation remains challenging. The classic
  Johnson--Neyman (JN) technique (Johnson and Neyman 1936) addresses this
  challenge for two-way interactions by identifying the regions of a
  moderator's range over which a focal effect is and is not statistically
  significant. The \pkg{int3ract} package for \proglang{R} implements the JN
  technique, extends it to three-way interactions over a two-dimensional
  moderator grid (JN3), and extends both to Bayesian models by working on
  posterior draws rather than point estimates. The package has a single entry
  point, the generic function \code{JN()}, which dispatches on the fitted
  object: models supplying point estimates and a covariance matrix are
  analyzed with Wald tests, objects supplying draws as conditional posterior
  distributions. Methods are provided for \code{lm()}/\code{glm()},
  \pkg{lme4}, \pkg{lavaan}, \pkg{RSiena} and \pkg{multiSiena} objects, and
  for matrices of
  posterior draws; support for any further model class is added by writing a
  single \code{jn\_input()} method, so that estimators from basically any package can be supported. Results are classed objects whose \code{summary()}
  method reports the regions of significance directly and whose \code{plot()}
  method draws shaded confidence bands (two-way) or crosshatched heatmaps
  (three-way), each accompanied by a panel showing how much data support each
  part of the moderator range, or the same regions as fans of
  simple slopes of the outcome. The package is designed to
  encourage richer, region-specific reporting of interaction effects in place
  of the conventional single-slope spotlight approach.
}

\Keywords{interaction effects, Johnson-Neyman technique,\\
          moderation analysis, SAOM, Bayesian, \proglang{R}}
\Plainkeywords{interaction effects, Johnson-Neyman technique,\\
               moderation analysis, SAOM, Bayesian, R}

\Address{
  Robert W.~Krause\\
  E-mail: \email{robert.w.krause@mailbox.org}
}

\providecommand{\Sconcordance}[1]{}

\begin{document}


\section{Introduction} \label{sec:intro}

Multiplicative interaction terms allow researchers to examine whether the
relationship between a predictor $X$ and an outcome $Y$ depends on the value
of one or more additional variables---the so-called moderators. Interpreting
such interactions by inspecting the sign and significance of the interaction
coefficient alone is, however, insufficient and potentially misleading
\citep{Bauer+Curran:2005,Preacher+Curran+Bauer:2006}. The key insight behind
the Johnson--Neyman (JN) technique is that, instead of evaluating the focal
effect at a small number of selected moderator values (often referred to as
``spotlight'' or ``simple-slope'' analysis), one can derive the \emph{entire}
region of the moderator's distribution over which the focal effect is
statistically distinguishable from zero
\citep{Johnson+Neyman:1936,Potthoff:1964}.

Originally proposed in the context of the analysis of covariance
\citep{Johnson+Neyman:1936}, the JN technique was later adapted for regression
\citep{Potthoff:1964} and updated by \citet{Bauer+Curran:2005}, whose
formulation is closest to what \pkg{int3ract} implements. The contribution of
the package is threefold.

\begin{enumerate}
  \item \textbf{Three-way interactions.} The package provides a complete
        implementation of the JN technique for three-way multiplicative
        interactions, referred to here as JN3. Because the significance region is then two-dimensional, it is
        rendered as a heatmap over the moderator grid rather than as a
        single-axis ribbon plot.

  \item \textbf{One interface, many model classes.} \code{JN()} is a generic
        that dispatches on the fitted object, with methods for \code{lm} and
        \code{glm}, \code{lmerMod} and \code{glmerMod}, \code{lavaan},
        \code{sienaFit}, \code{sienaBayesFit} and matrices of posterior
        draws. Support for further model classes requires a single
        \code{jn_input()} method and no change to the package itself
        (Section~\ref{sec:design}).

  \item \textbf{Bayesian models.} Where a model supplies posterior draws rather than
        point estimates, the same interface produces conditional posterior
        distributions and Bayesian $p$~values instead of Wald tests. Whether
        an analysis is frequentist or Bayesian is thus a property of the input,
        not a choice of function.
\end{enumerate}

Two further features address recurring difficulties in applied use. The
\code{summary()} method reports the regions of significance directly, rather
than leaving users to read them off a grid of conditional estimates. Additionally, every figure
shows the empirical distribution of the moderator to show how much data support the inference.

The remainder of the paper is organized as follows.
Section~\ref{sec:software} reviews related software. Section~\ref{sec:stat}
develops the statistical background of the JN technique and its three-way
extension. Section~\ref{sec:design} describes the design of the package.
Section~\ref{sec:freq} illustrates the frequentist application and
Section~\ref{sec:bayes} the Bayesian one. Section~\ref{sec:saom} covers
stochastic actor-oriented models (SAOMs). 
Section~\ref{sec:graphics} discusses graphical customization and
Section~\ref{sec:conc} concludes.


\section{Related software} \label{sec:software}

Visualizing what a fitted model implies, rather than reading its coefficient
table, is a well-established concern of this journal. \citet{Fox:2003}
introduced effect displays for generalized linear models in the \pkg{effects}
package, extended to multinomial and proportional-odds logit models by
\citet{Fox+Hong:2009}; \citet{Breheny+Burchett:2017} presented \pkg{visreg} for
conditional and contrast plots of arbitrary regression fits; and
\citet{ArelBundock+Greifer+Heiss:2024} described \pkg{marginaleffects}, which
computes predictions, comparisons and slopes---including their standard errors
by the delta method---for a very large set of model classes. \pkg{emmeans}
\citep{Lenth:2026} and \pkg{ggeffects} \citep{Ludecke:2018} serve the same
purpose from the estimated-marginal-means side. \pkg{int3ract} shares the
general aim of these packages and, in the two-way Wald case, the underlying
delta-method machinery. What distinguishes it is the object of interest: these
packages evaluate an effect \emph{at} chosen values of the moderator, whereas
the JN technique solves for the values at which the effect changes status.

Among packages targeting interactions specifically, \pkg{interactions}
\citep{Long:2019} and \pkg{jtools} \citep{Long:2020} provide the most complete
two-way JN implementation in \proglang{R}, including the boundary values and
the shaded plot. \citet{Lin:2020} reviews the available tools for producing
two-way JN figures and reaches the same conclusion for the wider software
landscape: implementations are plentiful for two-way interactions in
frequentist linear models, and absent elsewhere.

Outside \proglang{R} the same pattern holds. The most widely used implementation of the JN technique
in any environment is Hayes' \pkg{PROCESS} macro for \proglang{SPSS},
\proglang{SAS} and \proglang{R} \citep{Hayes:2022}, whose lineage runs back to
the \pkg{MODPROBE} macros of \citet{Hayes+Matthes:2009}; \citet{Lin:2020} rates
it among the best available. \proglang{Mplus} \citep{Mplus} reports regions of
significance for two-way interactions as part of its \code{LOOP} plotting
facilities. In \proglang{Stata} \citep{Stata} the established idiom is
\code{margins} together with \code{marginsplot}, which, like the effect-display
packages above, evaluates conditional effects at chosen moderator values rather
than solving for the values at which significance changes. In \proglang{Python}
the closest equivalent is again \pkg{marginaleffects}
\citep{ArelBundock+Greifer+Heiss:2024}.

What none of these environments provides is what \pkg{int3ract} introduces. To
the author's knowledge no package in \proglang{R} or elsewhere offers (i) JN
regions for three-way
interactions, where the region is a subset of a plane rather than an interval;
(ii) the JN technique applied to posterior draws, so that the same reasoning
extends to Bayesian estimation; or (iii) support for stochastic actor-oriented
models, where interactions between network statistics and actor attributes are
both common and hard to interpret, as well as model an easily extendable model-agnostic interface. The packages listed above cover
two-way frequentist interactions more thoroughly than \pkg{int3ract} attempts
to---in particular for factor-by-numeric interactions and for models
\pkg{int3ract} has no extractor for---and are the better choice for that case
when the additional features are not needed.


\section{Statistical background} \label{sec:stat}

\subsection{The two-way Johnson--Neyman technique}

Consider the linear model
\begin{equation} \label{eq:2way}
  Y_i \;=\; \beta_0 + \beta_1 X_i + \beta_2 M_i + \beta_3 X_i M_i + \varepsilon_i,
\end{equation}
where $X$ is the focal predictor, $M$ is the moderator, and $\beta_3$ is the
interaction coefficient. The simple slope of $Y$ with respect to $X$ at a given
value $m$ of the moderator is
\begin{equation} \label{eq:simpleslope}
  \hat{\theta}_X(m) \;=\; \hat{\beta}_1 + \hat{\beta}_3\,m.
\end{equation}
Its sampling variance follows from the delta method:
\begin{equation} \label{eq:var2way}
  \widehat{\mathrm{Var}}\!\bigl[\hat{\theta}_X(m)\bigr]
    \;=\; \widehat{\mathrm{Var}}(\hat{\beta}_1)
        + 2m\,\widehat{\mathrm{Cov}}(\hat{\beta}_1,\hat{\beta}_3)
        + m^{2}\,\widehat{\mathrm{Var}}(\hat{\beta}_3).
\end{equation}

The two-way JN regions are the set of moderator values $m$ for which the
two-sided $z$~test
\begin{equation} \label{eq:ztest}
  z(m) \;=\; \frac{\hat{\theta}_X(m)}
                  {\sqrt{\widehat{\mathrm{Var}}
                         \bigl[\hat{\theta}_X(m)\bigr]}}
\end{equation}
satisfies $|z(m)| > z_{\alpha/2}$, where $z_{\alpha/2}$ is the
$(\alpha/2)$-upper quantile of the standard normal distribution. Because both
$\hat{\theta}_X(m)^2$ and $\widehat{\mathrm{Var}}[\hat{\theta}_X(m)]$ are
quadratic in $m$, this condition is equivalent to the positivity of a single
quadratic,
\begin{equation} \label{eq:jnquad}
  \bigl(\hat{\beta}_3^{2} - z_{\alpha/2}^{2}\widehat{\mathrm{Var}}(\hat{\beta}_3)\bigr) m^{2}
  + 2\bigl(\hat{\beta}_1\hat{\beta}_3
      - z_{\alpha/2}^{2}\widehat{\mathrm{Cov}}(\hat{\beta}_1,\hat{\beta}_3)\bigr) m
  + \bigl(\hat{\beta}_1^{2} - z_{\alpha/2}^{2}\widehat{\mathrm{Var}}(\hat{\beta}_1)\bigr)
  \;>\; 0 ,
\end{equation}
whose roots are the JN boundaries. \pkg{int3ract} solves
Equation~\ref{eq:jnquad} directly, so the reported boundaries do not depend on
how finely the moderator range was gridded for plotting.

The symmetry of Equation~\ref{eq:2way} implies that the roles of $X$ and $M$
are interchangeable: the simple slope of $Y$ on $M$ moderated by $X$ is
$\hat{\beta}_2 + \hat{\beta}_3\,x$, with an analogous variance formula.
\pkg{int3ract} therefore always produces \emph{two} JN analyses per two-way
interaction, and three per three-way interaction---one for each variable acting
as the focal predictor.

\subsection{The three-way extension} \label{sec:threeway}

When a third variable $W$ enters as an additional moderator, the full three-way
interaction model is
\begin{equation} \label{eq:3way}
  \begin{aligned}
    Y_i \;=\; & \beta_0 + \beta_1 X_i + \beta_2 M_i + \beta_3 W_i \\
              & + \beta_4 X_i M_i + \beta_5 X_i W_i
                + \beta_6 M_i W_i + \beta_7 X_i M_i W_i
                + \varepsilon_i.
  \end{aligned}
\end{equation}
The conditional effect of $X$ given $M$ and $W$ is
\begin{equation} \label{eq:theta3way}
  \hat{\theta}_X(m,w) \;=\; \hat{\beta}_1 + \hat{\beta}_4 m
                          + \hat{\beta}_5 w + \hat{\beta}_7 mw,
\end{equation}
and the corresponding variance is obtained via the multivariate delta method:
\begin{equation} \label{eq:se3way}
  \begin{aligned}
  \widehat{\mathrm{Var}}\!\bigl[\hat{\theta}_X(m,w)\bigr]
    \;=\; & \widehat{\mathrm{Var}}_{\hat{\beta}_1}
          + m^{2}\widehat{\mathrm{Var}}_{\hat{\beta}_4}
          + w^{2}\widehat{\mathrm{Var}}_{\hat{\beta}_5}
          + (mw)^{2}\widehat{\mathrm{Var}}_{\hat{\beta}_7} \\
        & + 2m\,\widehat{\mathrm{Cov}}(\hat{\beta}_1,\hat{\beta}_4)
          + 2w\,\widehat{\mathrm{Cov}}(\hat{\beta}_1,\hat{\beta}_5)
          + 2mw\,\widehat{\mathrm{Cov}}(\hat{\beta}_1,\hat{\beta}_7) \\
        & + 2mw\,\widehat{\mathrm{Cov}}(\hat{\beta}_4,\hat{\beta}_5)
          + 2m^{2}w\,\widehat{\mathrm{Cov}}(\hat{\beta}_4,\hat{\beta}_7)
          + 2mw^{2}\,\widehat{\mathrm{Cov}}(\hat{\beta}_5,\hat{\beta}_7).
  \end{aligned}
\end{equation}
Because the significance region is now two-dimensional, it cannot be depicted
as an interval on a line. \pkg{int3ract} instead renders a heatmap of the
conditional point estimates $\hat{\theta}_X(m,w)$ using a diverging color
scale, and superimposes a crosshatch pattern over cells where the null
hypothesis $\theta_X = 0$ cannot be rejected at the chosen $\alpha$ (default
$\alpha = 0.05$). \code{summary()} reports the region along one moderator
conditional on selected values of the other, since a two-dimensional region has
no compact interval representation.

Such a slice is a two-way problem again. Holding $W$ at $w$,
Equation~\ref{eq:theta3way} reads
$\hat{\theta}_X(m \mid w) = (\hat{\beta}_1 + \hat{\beta}_5 w)
                          + (\hat{\beta}_4 + \hat{\beta}_7 w)\,m$,
linear in $m$ with coefficients that are linear combinations of the estimates,
and its variance from Equation~\ref{eq:se3way} is quadratic in $m$. The
boundaries of the slice are therefore the roots of Equation~\ref{eq:jnquad}
with $\hat{\beta}_1$ and $\hat{\beta}_3$ replaced by these two combinations and
their variances and covariance taken from the full covariance matrix. The
simple-slopes figures of Section~\ref{sec:freq} use this to draw three-way
regions slice by slice with exact boundaries.

\subsection{Bayesian analogue}

For Bayesian models there is no single point estimate or analytic covariance
matrix, and the delta method is neither needed nor applicable. Instead, the
conditional effect is computed at every draw. Writing $\theta^{(s)}$ for the
$s$th draw of a parameter, the two-way case gives
\begin{equation} \label{eq:bayes2}
  \tilde{\theta}_X^{(s)}(m)
    \;=\; \theta^{(s)}_{X} + \theta^{(s)}_{XM} \cdot m ,
\end{equation}
and the three-way case the analogue of Equation~\ref{eq:theta3way}, yielding a
full conditional posterior distribution at each moderator value. These are
summarized by the posterior mean, the posterior standard deviation, and the
Bayesian $p$~value
$\Pr(\tilde{\theta}_X > 0 \mid \text{data}, \text{model}, \text{prior})$. A
conditional effect counts as ``significant'' for the shading where that probability falls outside a
pair of thresholds, by default $(\alpha/2, 1-\alpha/2)$.


\section{Package design} \label{sec:design}

\subsection{Dispatch and the two carriers}

The Johnson--Neyman computation needs remarkably little from a fitted model:
either a handful of coefficients together with their covariance matrix, or a
matrix of draws of those same parameters. Everything else---the estimator, the
likelihood, the link function, the sampling scheme---is irrelevant once those
quantities are in hand.

The generic \code{jn_input()} turns a fitted object into one of exactly two
normalized carriers:

\begin{itemize}
  \item \code{jn_wald(coefficients, vcov, labels, ranges, support)}, holding
        point estimates and their covariance matrix, which leads to the Wald
        $z$~tests of Equation~\ref{eq:ztest}; and
  \item \code{jn_posterior(draws, labels, ranges, support)}, holding a matrix
        of draws, which leads to the conditional posteriors of
        Equation~\ref{eq:bayes2}.
\end{itemize}

Internally, parameters are stored in a canonical order so that
everything downstream is independent of how the original model named or ordered
its terms: $(\theta_1, \theta_2, \theta_{12})$ for two-way and $(\theta_1,
\theta_2, \theta_3, \theta_{12}, \theta_{13}, \theta_{23}, \theta_{123})$ for
three-way interactions. The user-facing generic \code{JN()} calls 
\code{jn_input()},
computes over the moderator grid, and returns a classed result; Figure~\ref{fig:design} shows the chain that allows all models to share one interface.

\begin{figure}[t!]
\centering
\begin{tikzpicture}[
  node distance = 5mm and 7mm,
  box/.style = {draw, rounded corners = 2pt, align = center,
                font = \small\ttfamily, inner sep = 4pt},
  lbl/.style = {align = center, font = \small},
  >= {Stealth[length = 5pt]}]

\node[box] (fit) {lm, glm, lmerMod,\\ lavaan, sienaFit,\\ mcmc, sienaBayesFit, \ldots};
\node[lbl, above = 1mm of fit] {fitted object};

\node[box, right = of fit] (input) {jn\_input()};
\node[lbl, above = 1mm of input] {method for its class};

\node[box, right = of input, yshift = 8mm] (wald) {jn\_wald};
\node[box, right = of input, yshift = -8mm] (post) {jn\_posterior};

\node[box, right = 12mm of wald, yshift = -8mm] (jn) {JN};
\node[lbl, above = 1mm of jn] {classed result};

\node[box, right = of jn] (meth) {print()\\ summary()\\ plot()\\ as.data.frame()};

\draw[->] (fit)   -- (input);
\draw[->] (input) -- (wald);
\draw[->] (input) -- (post);
\draw[->] (wald)  -- node[lbl, above, sloped, font = \scriptsize] {
} (jn);
\draw[->] (post)  -- node[lbl, below, sloped, font = \scriptsize] {
} (jn);
\draw[->] (jn)    -- (meth);
\end{tikzpicture}
\caption{\label{fig:design}The dispatch chain. A fitted object is normalized by
its \code{jn\_input()} method into one of two carriers; which carrier is
returned determines whether the analysis proceeds by Wald tests or by
conditional posteriors. Everything to the right of the carriers is shared.}
\end{figure}

Table~\ref{tab:methods} lists the \code{jn_input()} methods shipped with the
package.

\begin{table}[t!]
\centering
\small
\begin{tabular}{l l l}
\toprule
Model class & Fitted by & Carrier \\
\midrule
\code{lm}, \code{glm}          & \code{stats::lm()}, \code{stats::glm()} & \code{jn_wald} \\
\code{lmerMod}, \code{glmerMod} & \pkg{lme4} & \code{jn_wald} \\
\code{lavaan}                  & \pkg{lavaan} & \code{jn_wald} \\
\code{sienaFit}                & \pkg{RSiena} & \code{jn_wald} \\
\code{matrix}, \code{data.frame} & any sampler & \code{jn_posterior} \\
\code{mcmc}, \code{mcmc.list}  & \pkg{coda}, \pkg{MCMCpack} & \code{jn_posterior} \\
\code{sienaBayesFit}, \code{multiSiena} & \pkg{multiSiena} & \code{jn_posterior} \\
\bottomrule
\end{tabular}
\caption{\label{tab:methods}Model classes supported out of the box, and the
carrier each is normalized into.}
\end{table}

\subsection{The returned object and its methods}

\code{JN()} returns an object of class \code{JN}, carrying the moderator grid,
the conditional estimates and their uncertainty, the observed moderator values
where these are available, and the input carrier. Notably it does \emph{not}
carry the figures: these are built on demand by \code{plot()}, which keeps the
result small and lets the same analysis be drawn several ways.
Table~\ref{tab:jnmethods} lists the available methods. Analyses that produce
one result per group. Multi-level results from \pkg{lme4} with \code{fixed_only = FALSE} and
\pkg{multiSiena} models with \code{hyper_only = FALSE} return a
\code{JN_list} carrying the same set of methods for each group.

\begin{table}[t!]
\centering
\small
\begin{tabular}{l p{9.5cm}}
\toprule
Method & Returns \\
\midrule
\code{print()}          & a compact overview and the regions of significance \\
\code{summary()}        & the regions of significance in full, with the share
                          of observations inside each \\
\code{plot()}           & the figure for one focal variable, with data-density
                          panels; \code{type = "slopes"} draws the regions as
                          simple slopes of the outcome instead \\
\code{autoplot()}       & the same, for \pkg{ggplot2} users \\
\code{jn_plots()}       & every figure at once, named by focal variable \\
\code{jn_save()}        & writes all figures to disk \\
\code{jn_regions()}     & the regions of significance as a data frame \\
\code{as.data.frame()}  & the full grid of conditional effects \\
\code{coef()}, \code{vcov()} & the underlying parameters and their covariance \\
\bottomrule
\end{tabular}
\caption{\label{tab:jnmethods}Methods available for objects of class
\code{JN}.}
\end{table}

\subsection{Extending int3ract to a new model class} \label{sec:extend}

Supporting a model class the package has never heard of means writing one
\code{jn_input()} method for it. Consider a model fitted by variational
inference, where the posterior is approximated by a tractable variational
distribution rather than sampled. If that distribution is Gaussian, it is
summarized by a mean vector and a covariance matrix, and the natural carrier is
\code{jn_wald()}. For illustration, take an object of class \code{"vbfit"} with
those two components:
\begin{Schunk}
\begin{Sinput}
R> library("int3ract")
R> set.seed(1402)
R> dat <- data.frame(x = rnorm(100), z = rnorm(100), w = rnorm(100))
R> dat$y <- dat$x + 0.5 * dat$z + -0.5 * dat$w +
+    0.5 * dat$x * dat$z * dat$w + rnorm(100, sd = 4)
R> mod2 <- lm(y ~ x * z, data = dat)
R> pars <- c("x", "z", "x:z")
R> vb_fit <- structure(list(mu = coef(mod2)[pars],
+                           Sigma = vcov(mod2)[pars, pars]),
+                      class = "vbfit")
\end{Sinput}
\end{Schunk}
The method extracts the relevant parameters in the canonical order and hands
them to the constructor:
\begin{Schunk}
\begin{Sinput}
R> jn_input.vbfit <- function(object, theta_1, theta_2, ...) {
+    pars <- c(theta_1, theta_2, paste(theta_1, theta_2, sep = ":"))
+    jn_wald(coefficients = object$mu[pars],
+            vcov         = object$Sigma[pars, pars],
+            labels       = c(theta_1, theta_2))
+  }
\end{Sinput}
\end{Schunk}
That is the entire extension. \code{JN()} and all of its methods now work on
the new class:
\begin{Schunk}
\begin{Sinput}
R> jn_vb <- JN(vb_fit, theta_1 = "x", theta_2 = "z",
+              theta_1_vals = c(-3, 3), theta_2_vals = c(-3, 3))
R> class(jn_vb)
\end{Sinput}
\begin{Soutput}
[1] "JN_wald" "JN_2way" "JN"     
\end{Soutput}
\end{Schunk}
A variational family that is not Gaussian is handled just as easily: rather
than reporting its mean and covariance, the method simulates from it and
returns \code{jn_posterior(draws, labels)}, at which point the analysis follows
the Bayesian route of Equation~\ref{eq:bayes2} instead. The same two-line
pattern covers models fitted by penalized likelihood, by bootstrap, or by
anything else that yields either a covariance matrix or a set of draws.


\section{Frequentist usage} \label{sec:freq}

\subsection[Two-way interaction with JN]%
           {Two-way interaction with \code{JN()}}

For an \code{lm} or \code{glm} object, \code{JN()} needs only the names of the
two variables involved: coefficients, covariance sub-matrix, interaction term
and moderator ranges are all resolved automatically from the fitted model.
\begin{Schunk}
\begin{Sinput}
R> jn2 <- JN(mod2, theta_1 = "x", theta_2 = "z")
\end{Sinput}
\end{Schunk}
Printing the result, or calling \code{summary()} on it, reports the regions of
significance directly:
\begin{Schunk}
\begin{Sinput}
R> summary(jn2)
\end{Sinput}
\begin{Soutput}
Johnson-Neyman analysis (two-way, Wald z tests, alpha = 0.05)
Variables: x, z

Evaluated over
  x   [-2.622, 3.116]   1000 values   (observed [-2.144, 2.637])
  z   [-2.016, 2.586]   1000 values   (observed [-1.632, 2.203])

Regions of significance

  x, moderated by z
    z from   -1.979 to    1.371   effect positive   (88.0% of observations)

  z, moderated by x
    no region of significance over the evaluated range
\end{Soutput}
\end{Schunk}
The effect of \code{x} is significantly positive for values of \code{z}
between -1.979 and
1.371, which covers
88\% of the observations;
the effect of \code{z} is nowhere distinguishable from zero over the range of
\code{x} that was evaluated. The lower boundary lies below the smallest
observed value of \code{z}, -1.632. It is found at all
only because \code{JN()} evaluates each moderator a little beyond its observed
values---by default 10\% of the range on either side, set by the \code{extend}
argument---and \code{summary()} reports the
observed range next to the evaluated one for this reason. Within the data,
then, the effect of \code{x} is positive for every value of \code{z} up to the
upper boundary. A region that reaches the edge of the evaluated range is marked
with an asterisk to show that evaluation has stopped at the upper limit.

\begin{figure}[t!]
\centering
\begin{subfigure}[t]{0.48\textwidth}
  \centering
\includegraphics{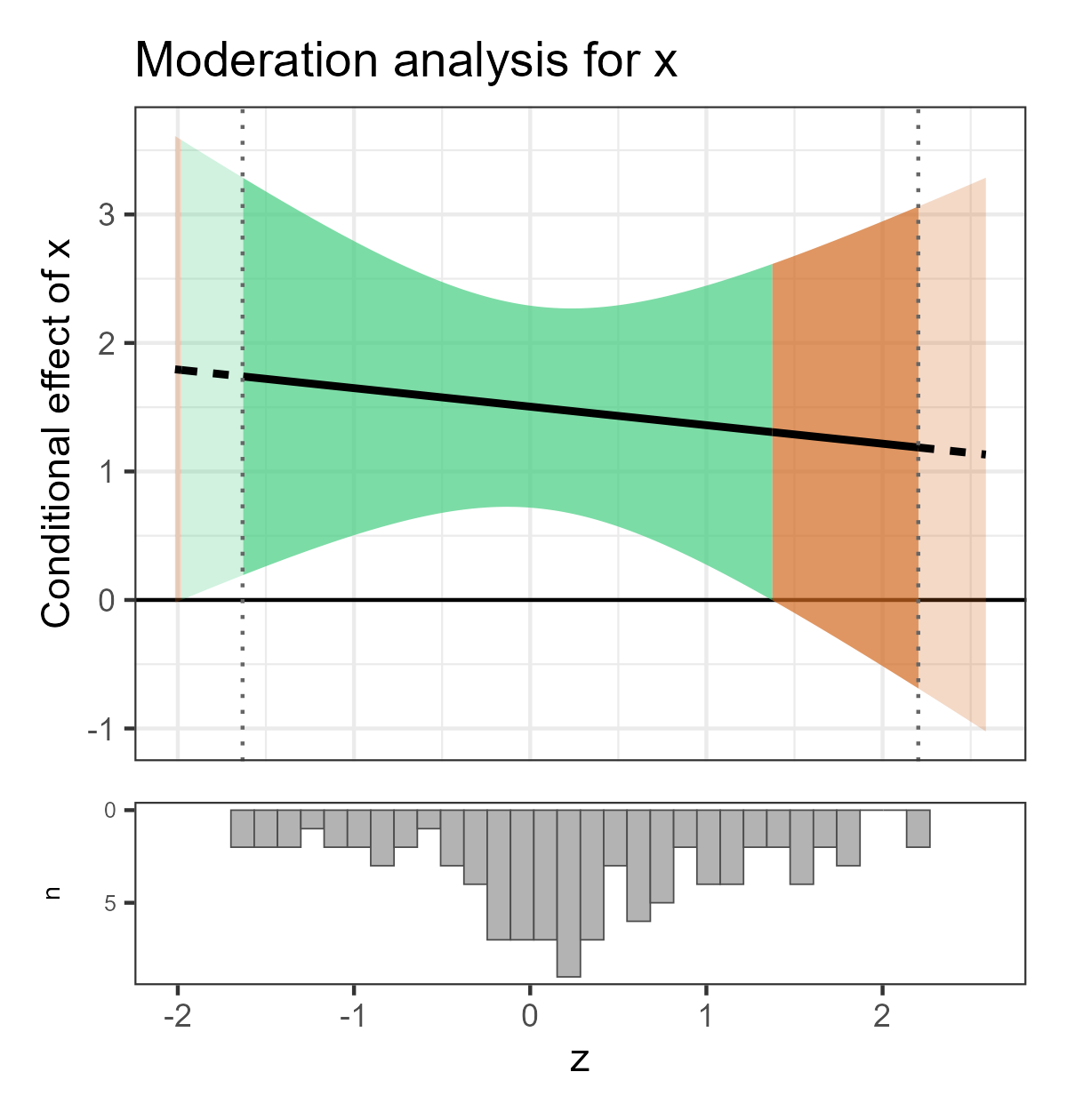}
  \caption{Conditional effect of \code{x} as a function of the moderator
           \code{z}.}
\end{subfigure}
\hfill
\begin{subfigure}[t]{0.48\textwidth}
  \centering
\includegraphics{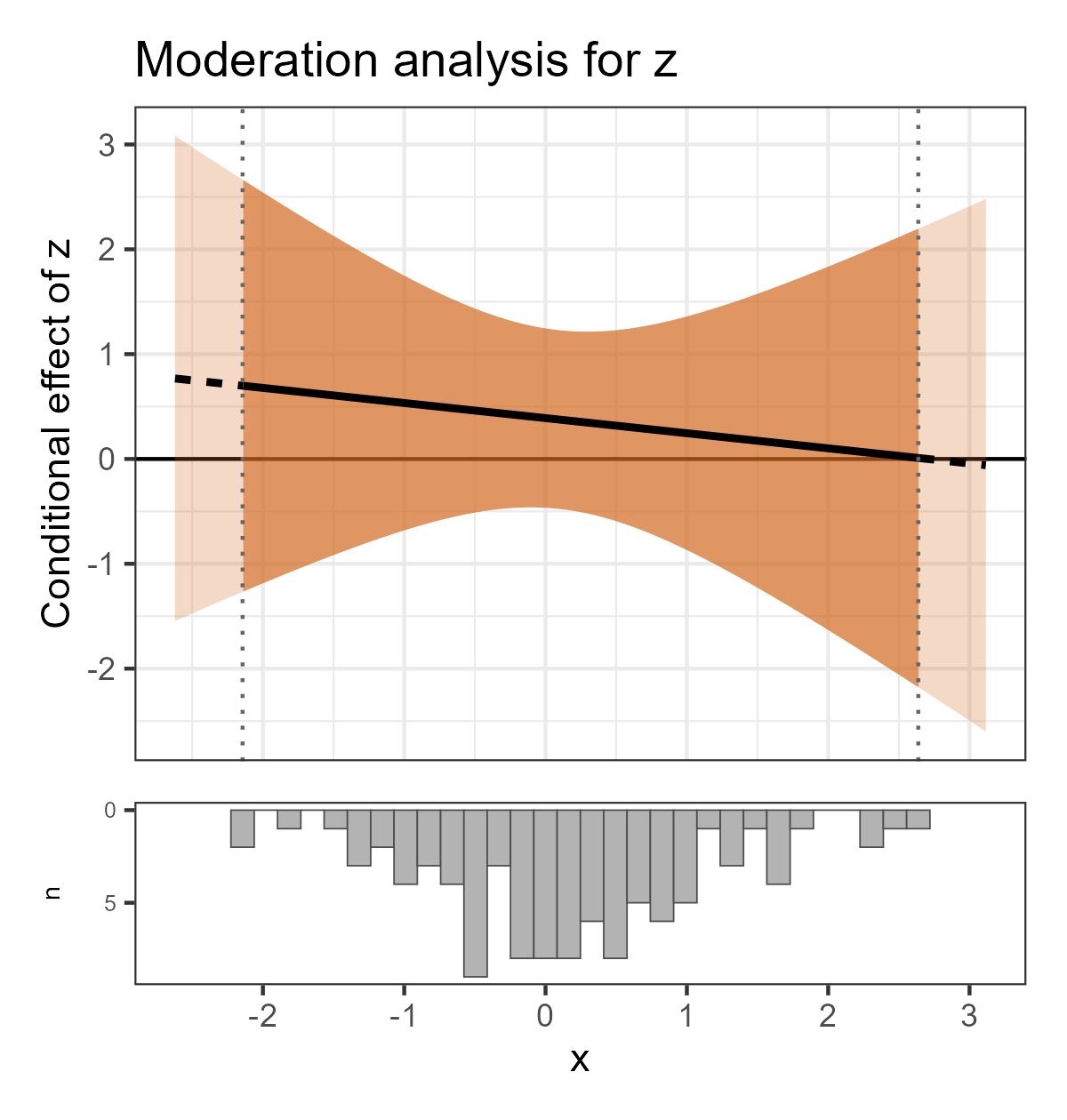}
  \caption{Conditional effect of \code{z} as a function of the moderator
           \code{x}.}
\end{subfigure}
\caption{Johnson--Neyman plots for the two-way interaction between \code{x}
         and \code{z} ($\alpha = 0.05$). The ribbon shows pointwise 95\%
         confidence bands for the conditional effect; green shading indicates
         regions where the effect is statistically significant, chocolate
         where it is not. The panel below each plot is a histogram of the
         observed moderator values, hanging downwards from the shared axis. Beyond the dotted lines the moderator
         was not observed; the band there is the margin added by
         \code{extend} and is drawn faint, with a dashed line.}
\label{fig:jn2}
\end{figure}

Because the roles of the two variables are interchangeable, there are two
figures to draw, and \code{plot(jn2)} draws both; \code{which} selects one of
them, as in \code{plot(jn2, which = "x")}. They are shown in
Figure~\ref{fig:jn2}.

\begin{figure}[t!]
\centering
\begin{subfigure}[t]{0.48\textwidth}
  \centering
\includegraphics{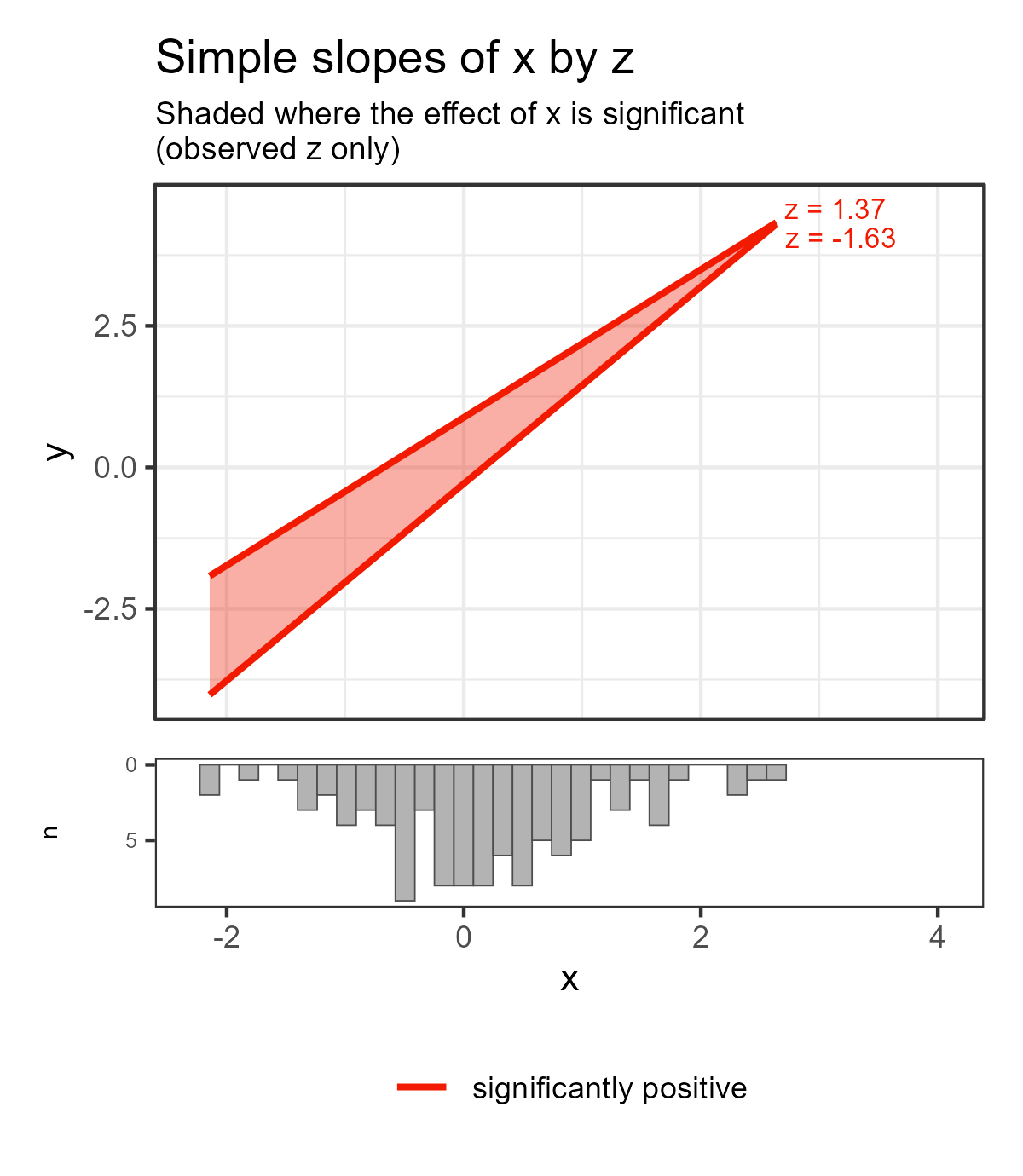}
  \caption{Simple slopes of \code{y} on \code{x} at the values of \code{z}
           bounding the region of significance.}
\end{subfigure}
\hfill
\begin{subfigure}[t]{0.48\textwidth}
  \centering
\includegraphics{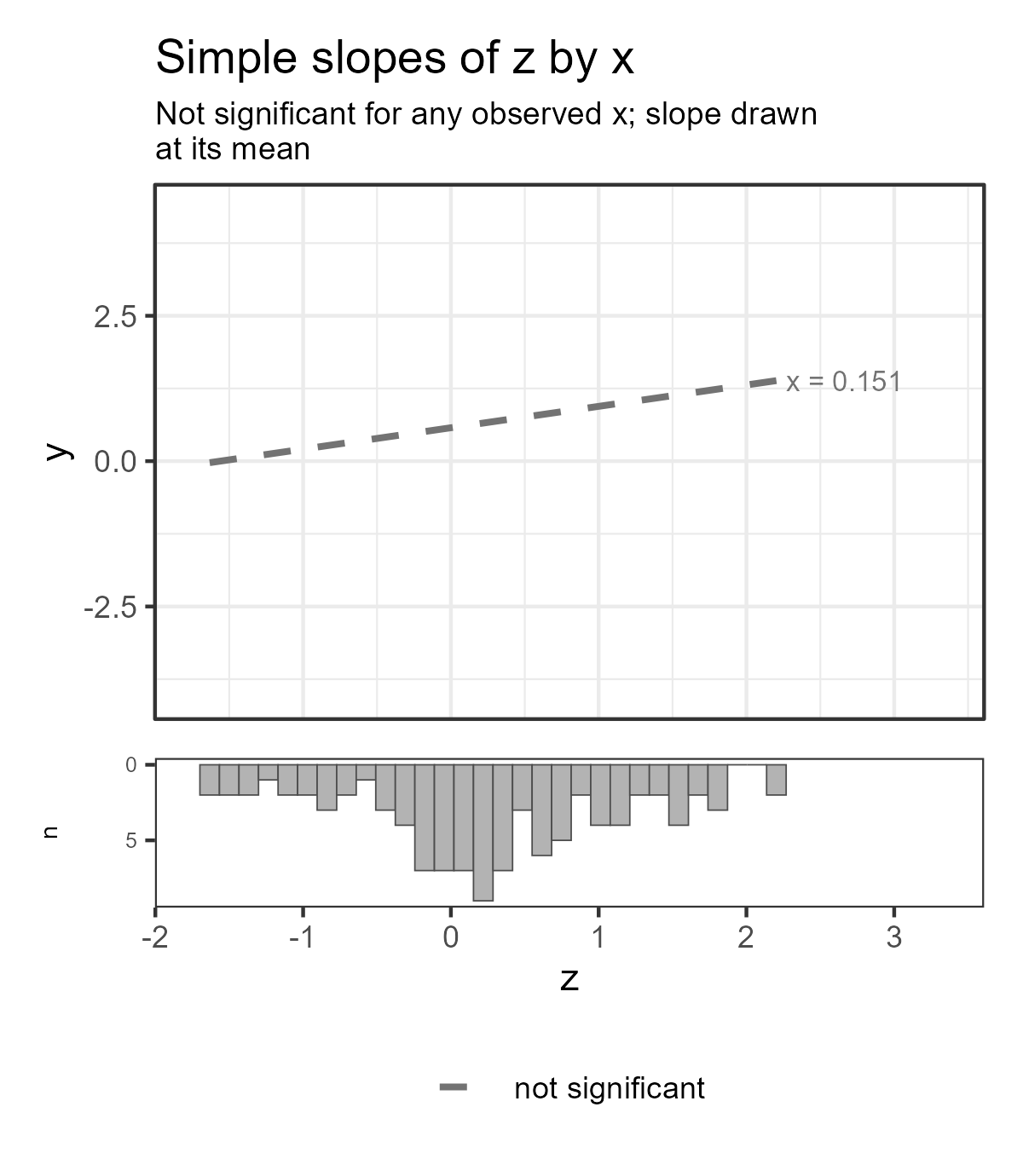}
  \caption{Simple slope of \code{y} on \code{z} at the mean of \code{x}; the
           effect of \code{z} is nowhere significant.}
\end{subfigure}
\caption{The regions of Figure~\ref{fig:jn2} drawn as simple slopes
         (\code{type = "slopes"}). Each slope is labeled with its moderator
         value; the shaded fan contains every significant simple slope. The
         panel below shows the distribution of the focal variable, which is
         now the horizontal axis.}
\label{fig:jn2-slopes}
\end{figure}

The JN figure is exact but abstract: its vertical axis is a slope, and readers
used to plots of simple slopes \citep{Preacher+Curran+Bauer:2006} must translate this to the classic $x$-$y$ axes. With
\code{type = "slopes"}, \code{plot()} draws the same regions in the outcome's
own terms (Figure~\ref{fig:jn2-slopes}). The focal variable is on the
horizontal axis and the model's prediction of the outcome on the vertical one,
on the scale of the linear predictor and with any further covariates held at
their means. A simple slope is drawn at each moderator value bounding a region
of significance, and the fan between the two bounding slopes is shaded red
where the focal effect is significantly positive and blue where it is
significantly negative, the two ends of the heatmap scale used below. Every
slope inside a shaded fan is significant. A region of one sign therefore
gives two slopes and regions of both signs four. A focal effect that is
nowhere significant gets a single dashed slope at the moderator's mean, as
for \code{z} here; one that is significant throughout gets the slopes at the
lowest and highest observed moderator values with a dotted third at the
moderator's most common value, which shows where most of the sample sits. The
regions are recomputed over the observed moderator values only, so a fan never
reaches into the margin added by \code{extend}. With
\code{jn_style(slope_shading = "density")} the fan is shaded by the density of
the moderator rather than uniformly, darkest where most observations lie.

The full grid of conditional estimates is available through
\code{as.data.frame()}, with one row per focal variable and moderator value;
its \code{extrapolated} column flags the moderator values lying beyond the
observed data:
\begin{Schunk}
\begin{Sinput}
R> head(as.data.frame(jn2), 3)
\end{Sinput}
\begin{Soutput}
  focal moderator mod_value estimate std.error statistic    p.value
1     x         z -2.015930 1.794906 0.9264632  1.937374 0.05269958
2     x         z -2.011324 1.794242 0.9247853  1.940172 0.05235884
3     x         z -2.006717 1.793578 0.9231081  1.942977 0.05201888
     conf.low conf.high significant extrapolated
1 -0.02092834  3.610741       FALSE         TRUE
2 -0.01830369  3.606788       FALSE         TRUE
3 -0.01568041  3.602837       FALSE         TRUE
\end{Soutput}
\end{Schunk}

\subsection[Three-way interaction with JN]%
           {Three-way interaction with \code{JN()}}

\begin{figure}[t!]
\centering
\begin{subfigure}[t]{0.48\textwidth}
  \centering
\includegraphics{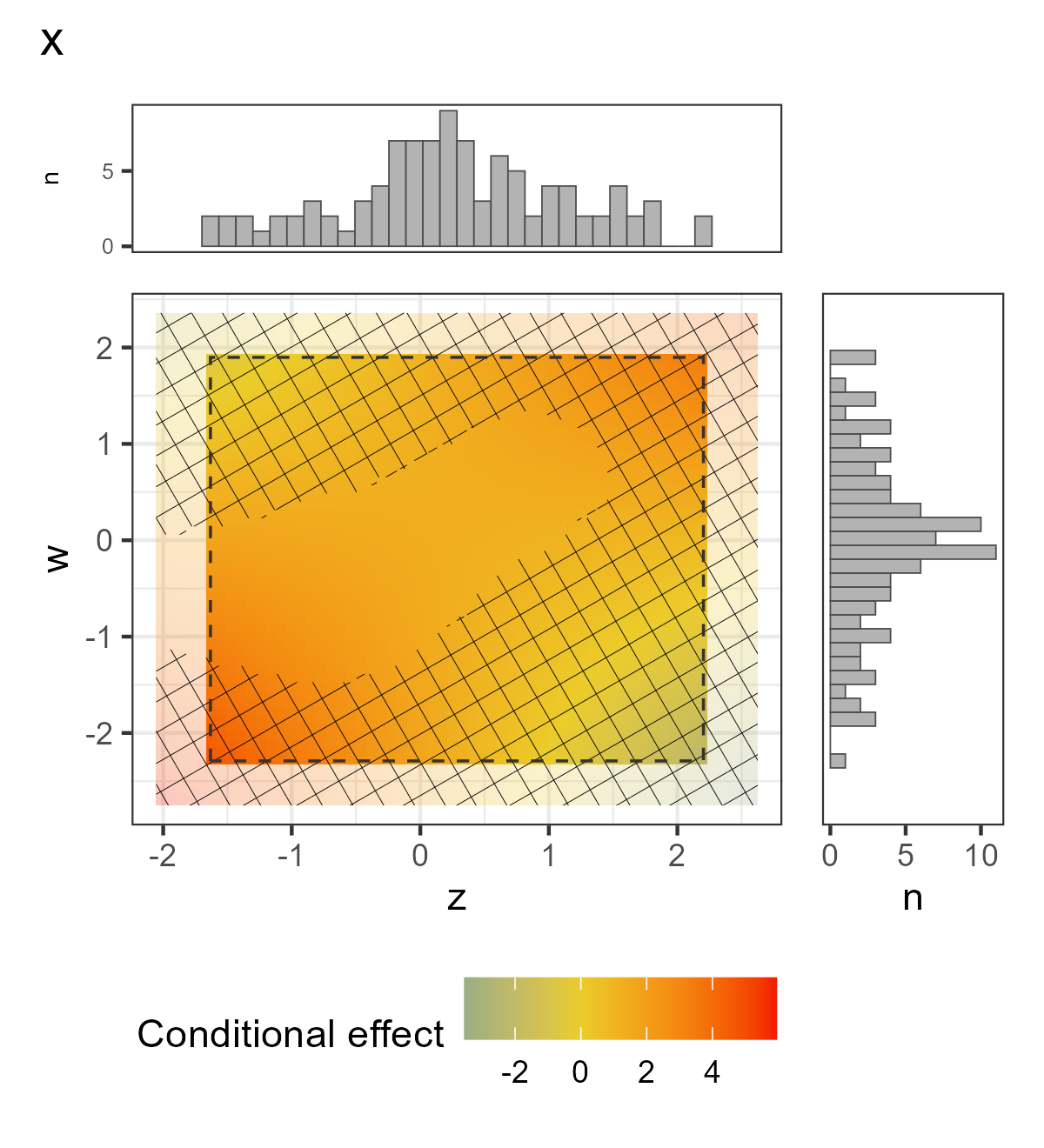}
  \caption{Conditional effect of \code{x} across the
           (\code{z}, \code{w}) moderator grid.}
\end{subfigure}
\hfill
\begin{subfigure}[t]{0.48\textwidth}
  \centering
\includegraphics{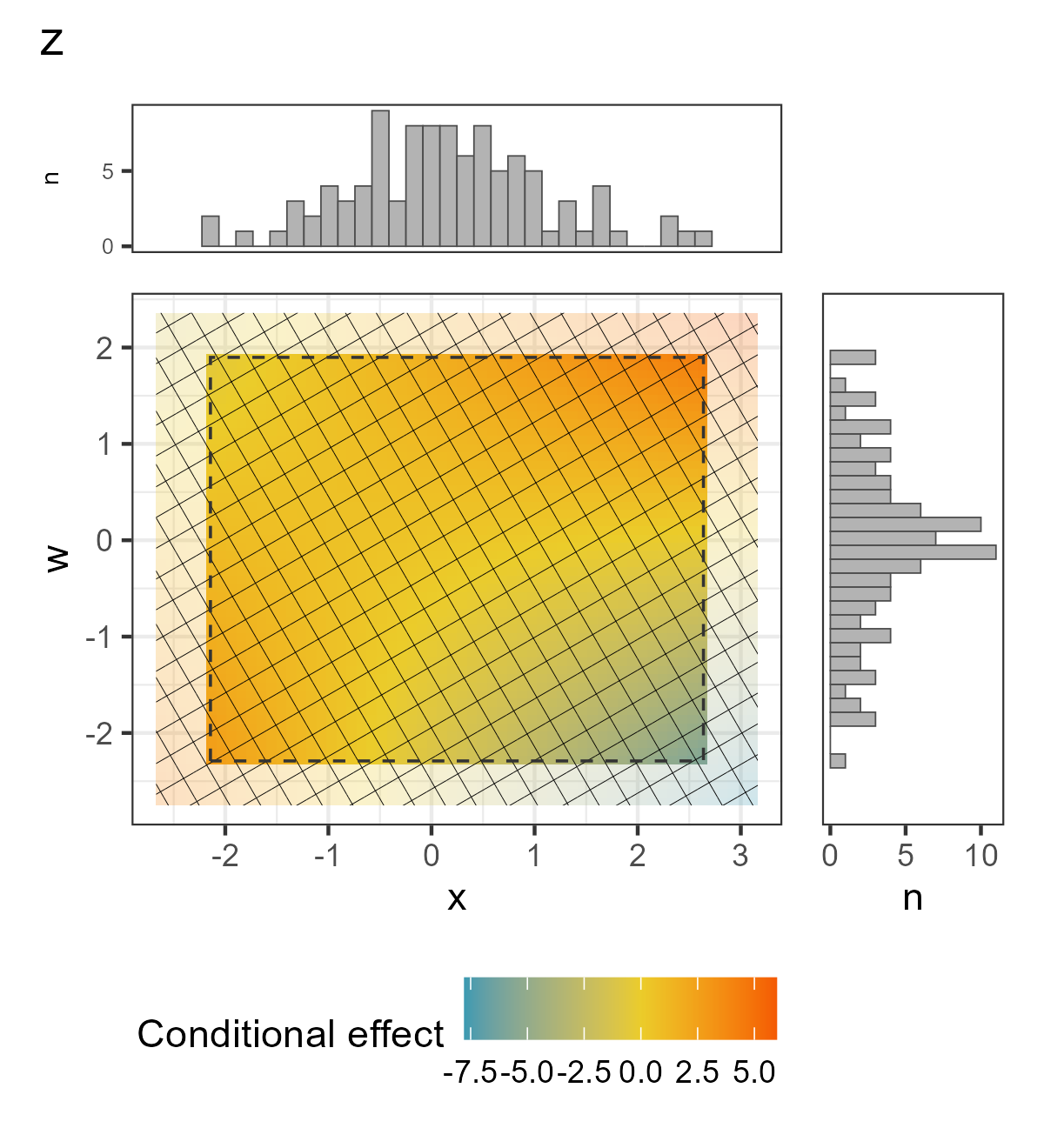}
  \caption{Conditional effect of \code{z} across the
           (\code{x}, \code{w}) moderator grid.}
\end{subfigure}

\medskip

\begin{subfigure}[t]{0.48\textwidth}
  \centering
\includegraphics{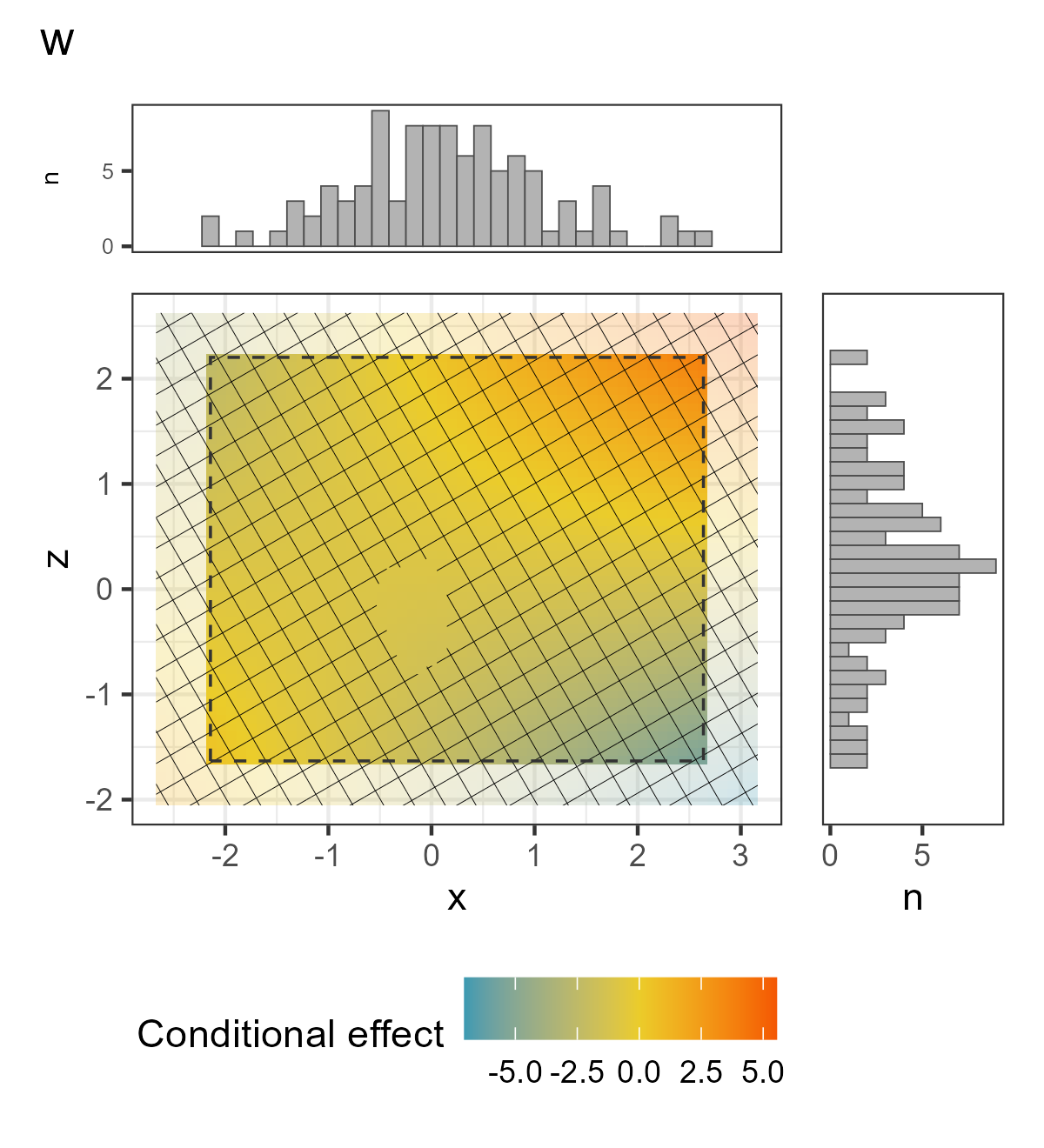}
  \caption{Conditional effect of \code{w} across the
           (\code{x}, \code{z}) moderator grid.}
\end{subfigure}
\caption{JN3 heatmaps for the three-way interaction among \code{x}, \code{z},
         and \code{w} (\code{range\_size = 60}, $\alpha = 0.05$). Each cell
         shows the conditional point estimate of the focal effect across the
         two-dimensional moderator grid; the diverging color scale runs from
         blue (negative) through yellow (zero) to red (positive). Crosshatched
         cells indicate regions where the conditional effect is not
         statistically significant. The histograms along the top and right
         edges show the marginal distributions of the two moderators. The
         dashed rectangle encloses the observed moderator values; the faint
         cells outside it are the margin added by \code{extend}.}
\label{fig:jn3}
\end{figure}

The syntax for three-way interactions differs only in that \code{theta_3} is
also supplied.
\begin{Schunk}
\begin{Sinput}
R> mod3 <- lm(y ~ x * z * w, data = dat)
R> jn3 <- JN(mod3, theta_1 = "x", theta_2 = "z", theta_3 = "w",
+            range_size = 60)
\end{Sinput}
\end{Schunk}
Because the region of significance is now two-dimensional, \code{summary()}
reports the interval along the second moderator at selected values of the
first---by default its quartiles. The values to condition on can be chosen with
the \code{at} argument.
\begin{Schunk}
\begin{Sinput}
R> summary(jn3, at = 0)
\end{Sinput}
\begin{Soutput}
Johnson-Neyman analysis (three-way, Wald z tests, alpha = 0.05)
Variables: x, z, w

Evaluated over
  x   [-2.622, 3.116]   60 values   (observed [-2.144, 2.637])
  z   [-2.016, 2.586]   60 values   (observed [-1.632, 2.203])
  w   [-2.709, 2.317]   60 values   (observed [-2.290, 1.898])

Regions of significance

  x, moderated by z and w
    at z =    0.012
      w from   -1.103 to    0.848   effect positive   (70.0% of w values)

  z, moderated by x and w
    at x =    0.004
      no region of significance over the evaluated range

  w, moderated by x and z
    at x =    0.004
      z from   -0.714 to    0.253   effect negative   (39.0% of z values)
\end{Soutput}
\end{Schunk}
The heatmaps are shown in Figure~\ref{fig:jn3}. Setting
\code{range_size = 60} produces a $60 \times 60$ grid for each of the three
heatmaps, that is $3 \times 3{,}600$ conditional effect evaluations. The
evaluations themselves are cheap; drawing that many tiles is not, and rendering
time in \pkg{ggplot2} grows quickly with the grid. It is therefore worth
exploring with smaller values (e.g., \code{range_size = 25}) and reserving finer
grids for publication figures. When every moderator range contains fewer
than eight distinct values, the point estimate is printed inside each cell;
otherwise the color gradient alone communicates magnitude. Moderator ranges are
taken from the model where possible, and must be supplied via
\code{theta_1_vals} and friends for \code{sienaFit} objects and for carriers
built by hand.

\begin{figure}[t!]
\centering
\includegraphics{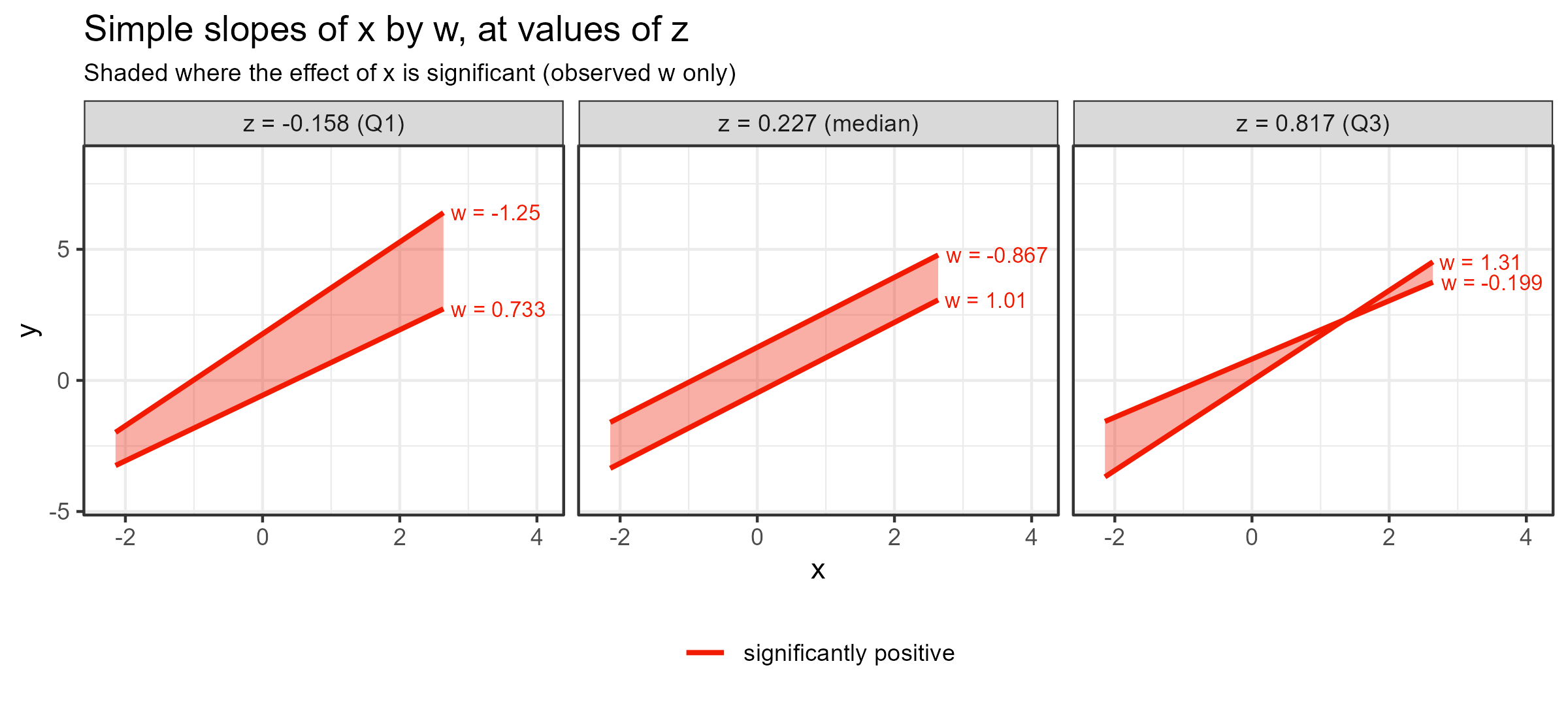}
\caption{Simple slopes of \code{y} on \code{x} in the three-way model, at the
         quartiles of \code{z} (one panel each). Within each panel the slopes
         bound the values of \code{w} for which the effect of \code{x} is
         significant, red for a positive and blue for a negative effect; a
         single dashed slope marks a panel in which it is nowhere
         significant.}
\label{fig:jn3-slopes}
\end{figure}

A three-way region has no pair of bounding values, so its simple-slopes figure
is drawn in slices (Figure~\ref{fig:jn3-slopes}): one panel per value of the
first moderator---its quartiles by default, or the values passed to
\code{plot()} as \code{at}---each showing the two-way figure of the focal
variable along the second moderator. Within a panel the region is a two-way
one, and for Wald analyses its boundaries are solved exactly as described in
Section~\ref{sec:threeway} rather than read off the grid. Reading across the
panels shows how the region along one moderator shifts, shrinks or changes sign
as the other moderator changes, which is the substance of a three-way
interaction.

\subsection{Binary moderators}

\begin{figure}[t!]
\centering
\begin{subfigure}[t]{0.48\textwidth}
  \centering
\includegraphics{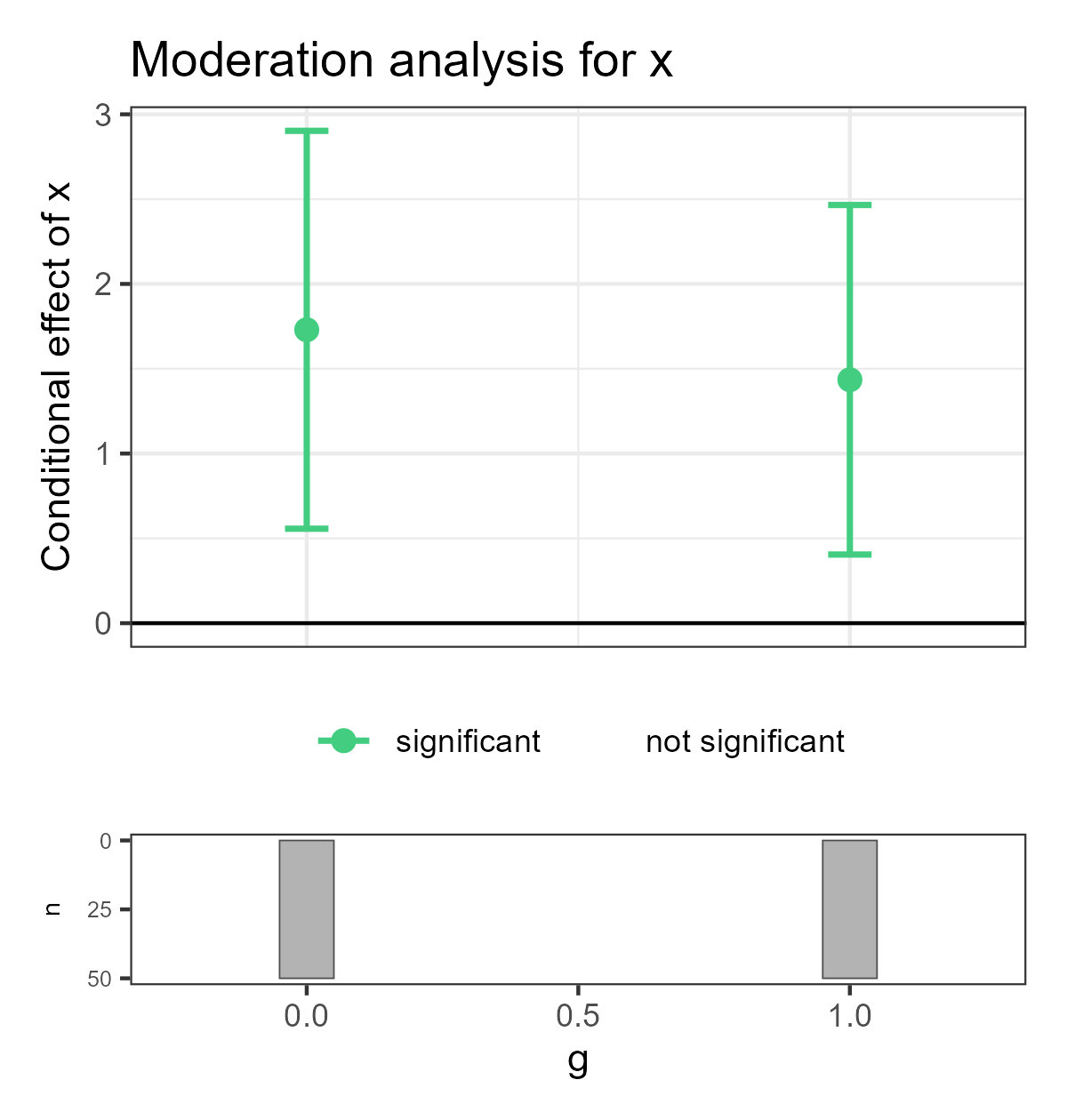}
  \caption{Conditional effect of \code{x} at the two values of \code{g}.}
\end{subfigure}
\hfill
\begin{subfigure}[t]{0.48\textwidth}
  \centering
\includegraphics{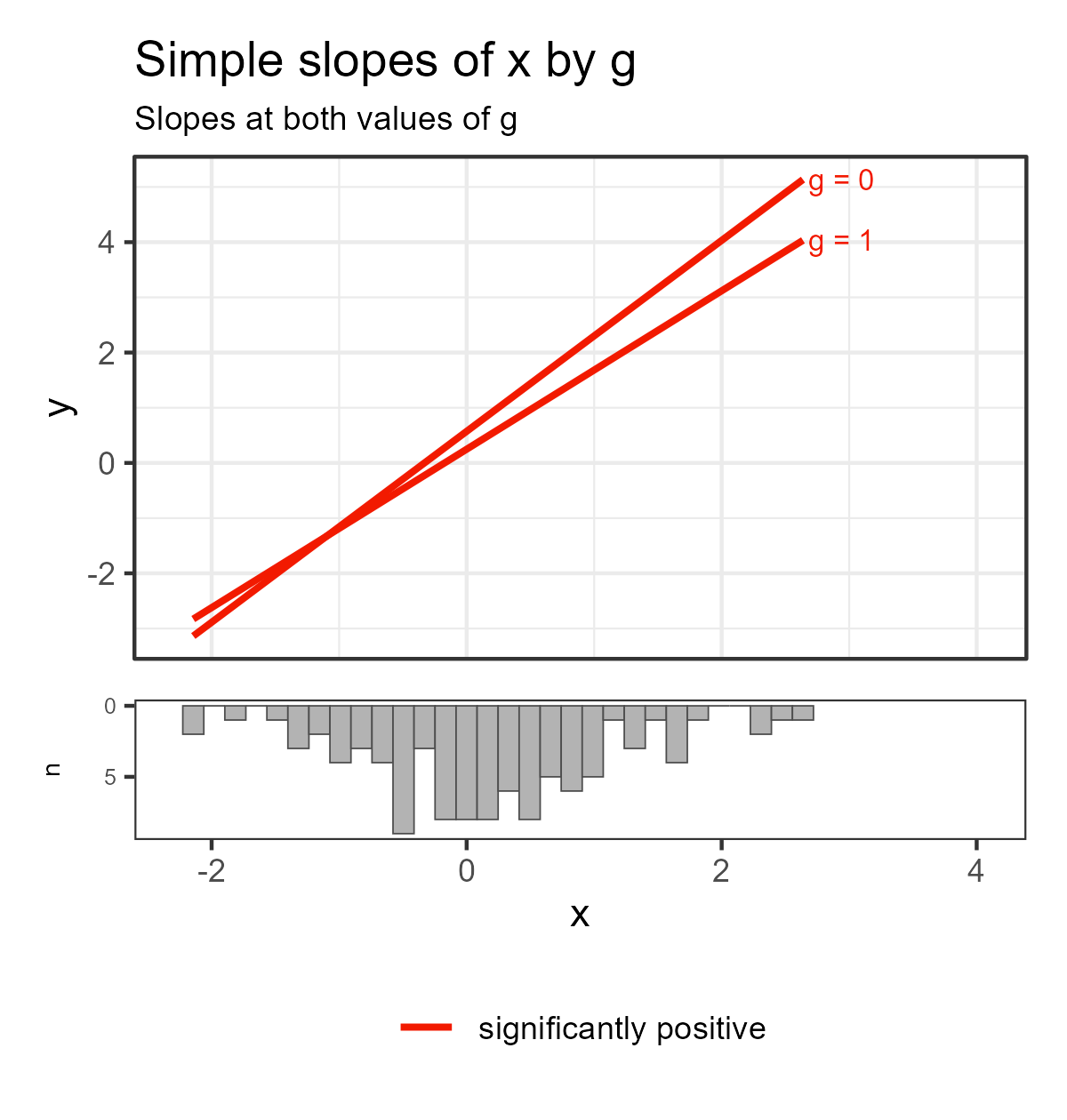}
  \caption{Simple slopes of \code{y} on \code{x} at the two values of
           \code{g}.}
\end{subfigure}
\caption{A binary moderator. The conditional effect is estimated at its two
         values only, and both simple slopes are drawn, each marked as
         significant or not.}
\label{fig:jn-binary}
\end{figure}

A moderator observed at exactly two values---a treatment indicator, say---is
recognized as binary and evaluated at those two values only; the \code{extend}
margin does not apply to it. The conditional effect is then shown as an
estimate with its confidence interval at each of the two values, and the
simple-slopes figure draws both slopes, each marked as significant or not, with
nothing shaded between them. For illustration, split \code{w} at zero:
\begin{Schunk}
\begin{Sinput}
R> dat$g <- as.numeric(dat$w > 0)
R> jn_bin <- JN(lm(y ~ x * g, data = dat), theta_1 = "x", theta_2 = "g")
\end{Sinput}
\end{Schunk}
The two figures for \code{x} are shown in Figure~\ref{fig:jn-binary}. In a
three-way interaction involving a binary variable, the heatmap over a grid
with only two rows would be misleading, and the figures are drawn as two
panels instead, one per value of the binary variable, each two-way figure
along the continuous moderator. Where the observed values are unknown---for
posterior draws or \code{sienaFit} objects without a \code{support}
argument---binary variables are named through \code{binary}, since a supplied
range such as \code{theta_2_vals = c(0, 1)} is otherwise read as a range.

\subsection[Mixed-effects models (lme4)]%
           {Mixed-effects models (\pkg{lme4})}

\code{JN()} has a method for \code{lmerMod} and \code{glmerMod} objects from
\pkg{lme4} \citep{lme4}. For fixed effects the interface is identical to that
for \code{lm}:
\begin{Schunk}
\begin{Sinput}
R> library("lme4")
R> lme_mod <- lmer(y ~ x * z + (x * z | group), data = dat)
R> jn_lme <- JN(lme_mod, theta_1 = "x", theta_2 = "z")
\end{Sinput}
\end{Schunk}
Setting \code{fixed_only = FALSE} additionally produces one analysis per level
of the grouping factor, using group-specific coefficients (fixed effects plus
conditional modes) and the fixed-effects covariance matrix. The result is a
\code{JN_list}; because inspecting dozens of figures interactively is
impractical in \proglang{R}, \code{jn_save()} writes them all to disk in one call. The grouping
variable defaults to the first random-effects term and can be set explicitly:
\begin{Schunk}
\begin{Sinput}
R> jn_groups <- JN(lme_mod, theta_1 = "x", theta_2 = "z",
+                  fixed_only = FALSE, group_var = "group")
R> jn_save(jn_groups, folder = "jn-figures")
\end{Sinput}
\end{Schunk}

\subsection[Structural equation and path models (lavaan)]%
           {Structural equation and path models (\pkg{lavaan})}

Interaction terms are as common in structural equation/path models as in regressions, but a
structural model differs in one respect that matters here: it usually has more
than one endogenous variable. The same product may enter several structural
equations with different coefficients, so ``the'' conditional effect is not
well defined until the equation is named. That is what the \code{outcome}
argument does, and it is the only addition the \pkg{lavaan} method requires to the
interface:
\begin{Schunk}
\begin{Sinput}
R> library("lavaan")
R> dat$y2 <- -dat$x + 0.3 * dat$z - 0.4 * dat$x * dat$z + rnorm(100, sd = 4)
R> sem_fit <- sem("y  ~ x + z + x:z
+                  y2 ~ x + z + x:z", data = dat)
R> JN(sem_fit, theta_1 = "x", theta_2 = "z", outcome = "y")
\end{Sinput}
\end{Schunk}
When \code{outcome} is omitted and exactly one endogenous variable is regressed
on all of the focal variables, that equation is used; when several are,
\code{JN()} reports which they are rather than choosing one. Latent outcomes
require nothing special and the moderators keep their data-density panels as long as they
are themselves observed. A model estimated from a sample covariance matrix
carries no raw data, so there the panels are absent and the moderator ranges
have to be supplied through \code{theta_1_vals} and \code{theta_2_vals}.

\pkg{lavaan} writes two-way products as \code{x:z} in its model syntax, and
those are located automatically. Three-way products it does not accept at all:
they must be constructed as variables in the data, under names of the user's
choosing, and those names passed through \code{theta_int_12} and its
companions---the same arguments that address effects positionally for
\code{sienaFit} objects.
\begin{Schunk}
\begin{Sinput}
R> dat$xz <- dat$x * dat$z
R> dat$xw <- dat$x * dat$w
R> dat$zw <- dat$z * dat$w
R> dat$xzw <- dat$x * dat$z * dat$w
R> sem3 <- sem("y ~ x + z + w + xz + xw + zw + xzw", data = dat)
R> JN(sem3, theta_1 = "x", theta_2 = "z", theta_3 = "w",
+     theta_int_12 = "xz", theta_int_13 = "xw",
+     theta_int_23 = "zw", theta_int_123 = "xzw")
\end{Sinput}
\end{Schunk}
Multiple-group fits are not supported; fitting the groups separately and calling \code{JN()} on each gives
the intended comparison. A \pkg{blavaan} is also not supported. Its
draws should be extracted and passed to \code{JN()} directly, as in
Section~\ref{sec:bayes}.

\subsection{Direct use with coefficient vectors}

When no \code{jn_input()} method exists and writing one is not worth the
trouble or possible (because no \code{classed} object is created by the analysis algorithm), the carrier can be constructed
directly and handed to \code{JN()}:
\begin{Schunk}
\begin{Sinput}
R> inp <- jn_wald(coefficients = my_coefs[c("x", "z", "x:z")],
+                 vcov = my_vcov[c("x", "z", "x:z"), c("x", "z", "x:z")],
+                 labels = c("x", "z"),
+                 ranges = list(c(-3, 3), c(-3, 3)),
+                 support = list(my_data$x, my_data$z))
R> JN(inp)
\end{Sinput}
\end{Schunk}
%


\section{Bayesian usage} \label{sec:bayes}

\subsection{Two-way interactions}

Objects carrying draws take the same route through \code{JN()}. A matrix of
posterior draws, an \code{mcmc} object from \pkg{coda} or \pkg{MCMCpack}, or a
\code{sienaBayesFit} object are all normalized to \code{jn_posterior} and
analyzed as conditional posteriors.

\begin{figure}[t!]
\centering
\begin{subfigure}[t]{0.48\textwidth}
  \centering
\includegraphics{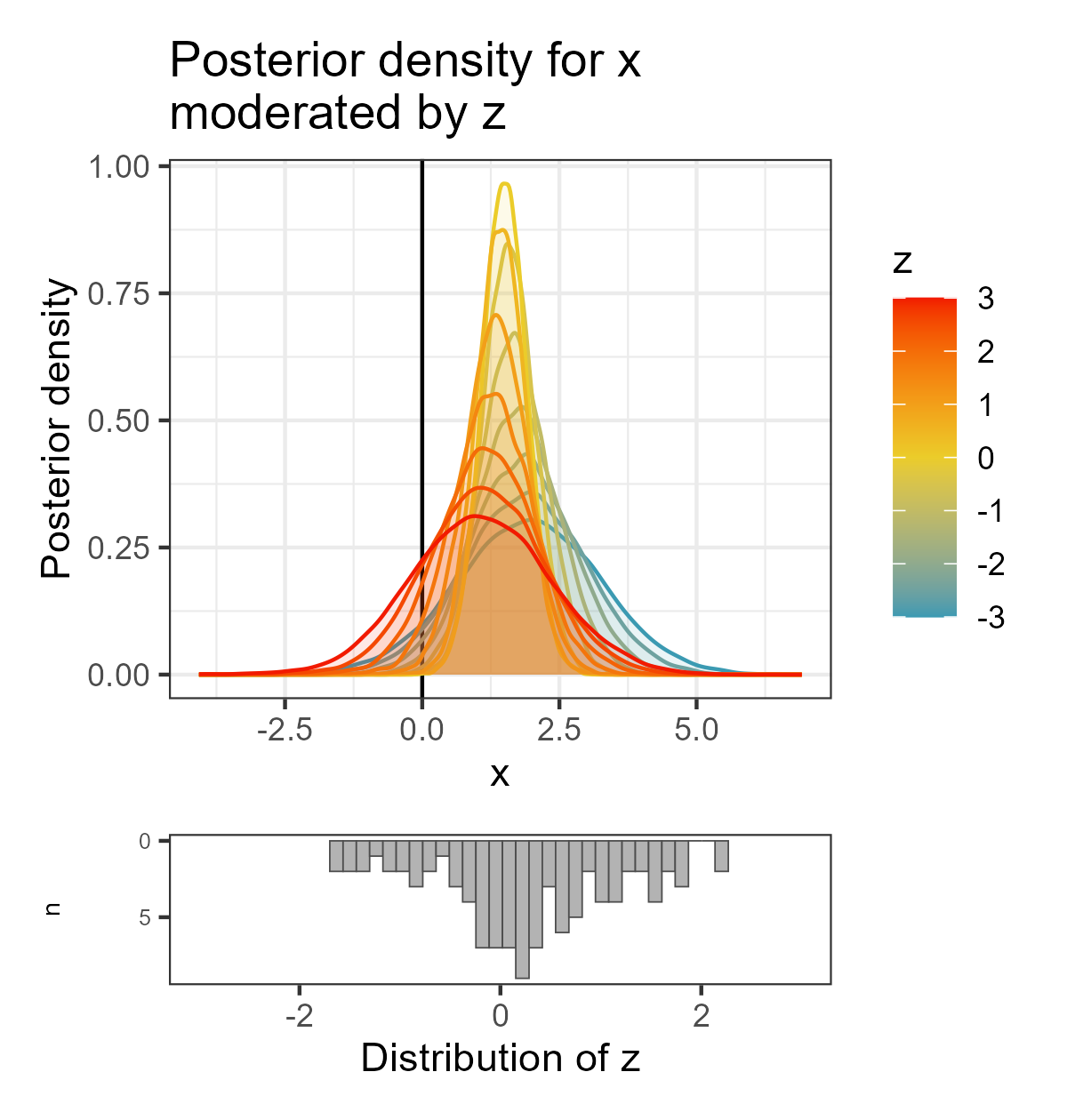}
  \caption{Conditional posteriors of the effect of \code{x} at each value of
           the moderator \code{z}.}
\end{subfigure}
\hfill
\begin{subfigure}[t]{0.48\textwidth}
  \centering
\includegraphics{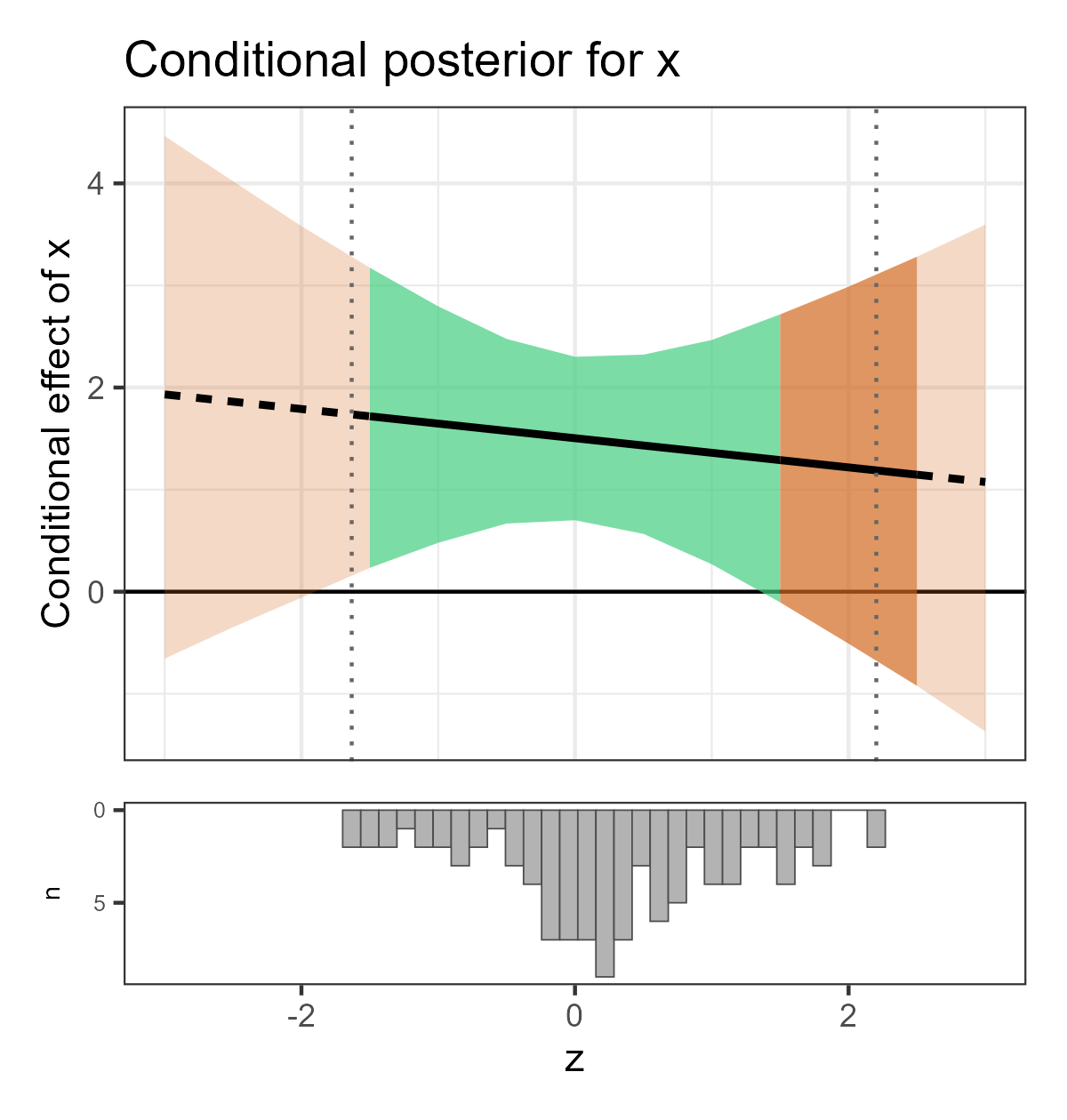}
  \caption{The same analysis drawn as a posterior mean with its 95\%
           credible band.}
\end{subfigure}
\caption{Two ways of drawing a two-way Bayesian JN analysis. On the left, one
         conditional posterior density per moderator value, colored by that
         value; the panel below shows the distribution of the moderator itself.
         On the right, \code{type = "band"} gives the posterior mean and
         credible band against the moderator, shaded by ``significance'', which is
         directly comparable to the frequentist figure.}
\label{fig:jn-bayes2}
\end{figure}

Parameters can be given by name---in which case the columns of the draw matrix
must be named accordingly, with interactions in the standard \proglang{R}
notation such as \code{x:z}, and the interaction columns are located
automatically---or by integer position. Unlike the frequentist case the
moderator values must always be supplied, because Bayesian fitting functions in
\proglang{R} do not consistently retain the data they were given. Each supplied
value produces one conditional posterior, so a long vector yields an
unreadable figure; a dozen values is usually plenty. Burn-in and thinning are
available through \code{burn_in} and \code{thin} (defaults: none and one).
\begin{Schunk}
\begin{Sinput}
R> library("MCMCpack")
R> mod_bayes2 <- MCMCregress(y ~ x * z, data = dat,
+                            burnin = 1000, mcmc = 10000, thin = 1,
+                            verbose = 0)
R> jn_bayes2 <- JN(mod_bayes2, theta_1 = "x", theta_2 = "z",
+                   theta_1_vals = seq(-3, 3, 0.5),
+                   theta_2_vals = seq(-3, 3, 0.5),
+                   support = list(dat$x, dat$z))
\end{Sinput}
\end{Schunk}
Note the \code{support} argument: since the posterior draws carry no data,
the observed moderator values are supplied explicitly so that the figures can
show them. The conditional posteriors are shown in
Figure~\ref{fig:jn-bayes2}.

\subsection{Three-way interactions}

\begin{figure}[t!]
\centering
\begin{subfigure}[t]{0.48\textwidth}
  \centering
\includegraphics{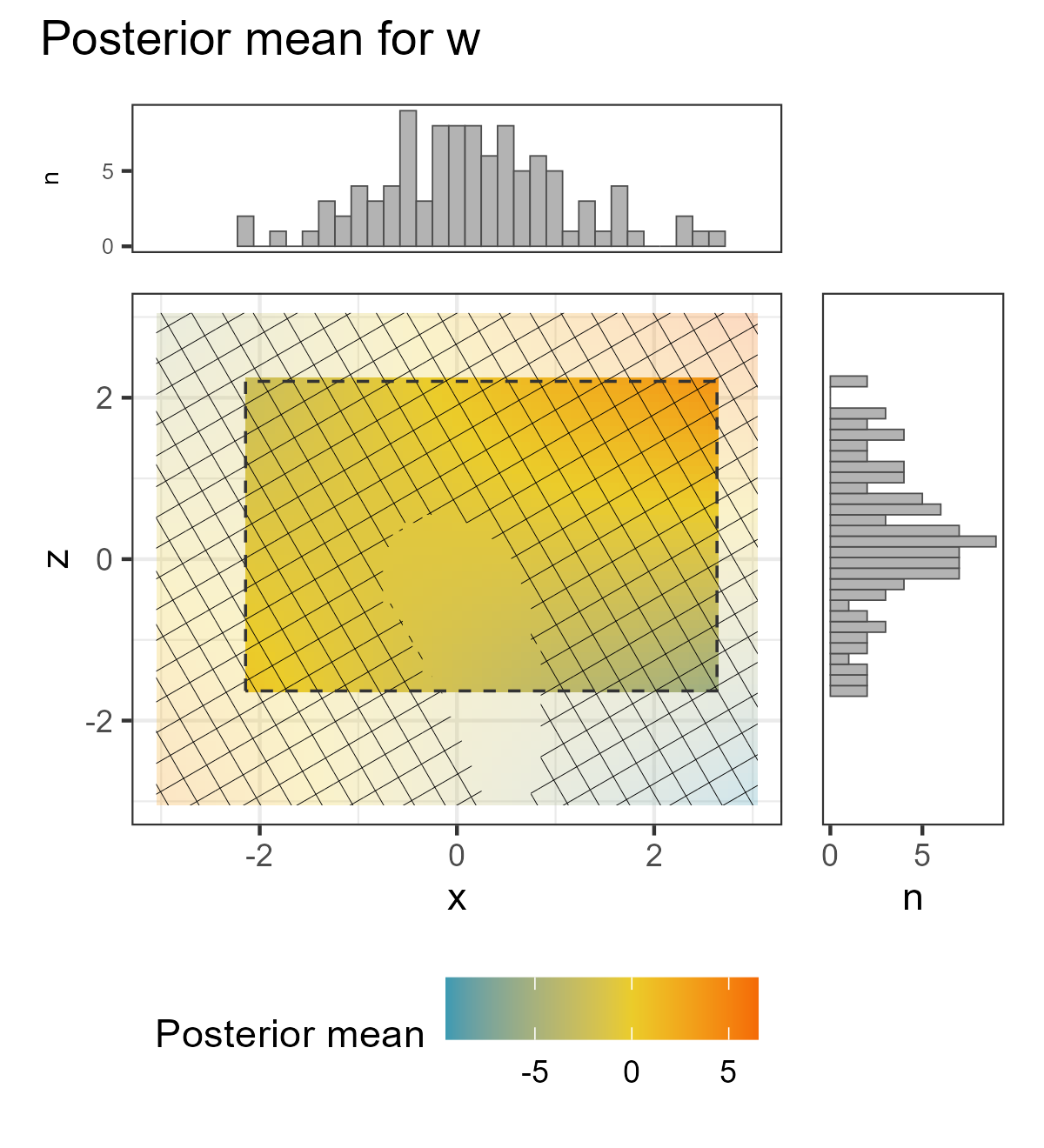}
  \caption{Conditional posterior means of the effect of \code{w} as a function
           of the moderators \code{x} and \code{z}.}
\end{subfigure}
\hfill
\begin{subfigure}[t]{0.48\textwidth}
  \centering
\includegraphics{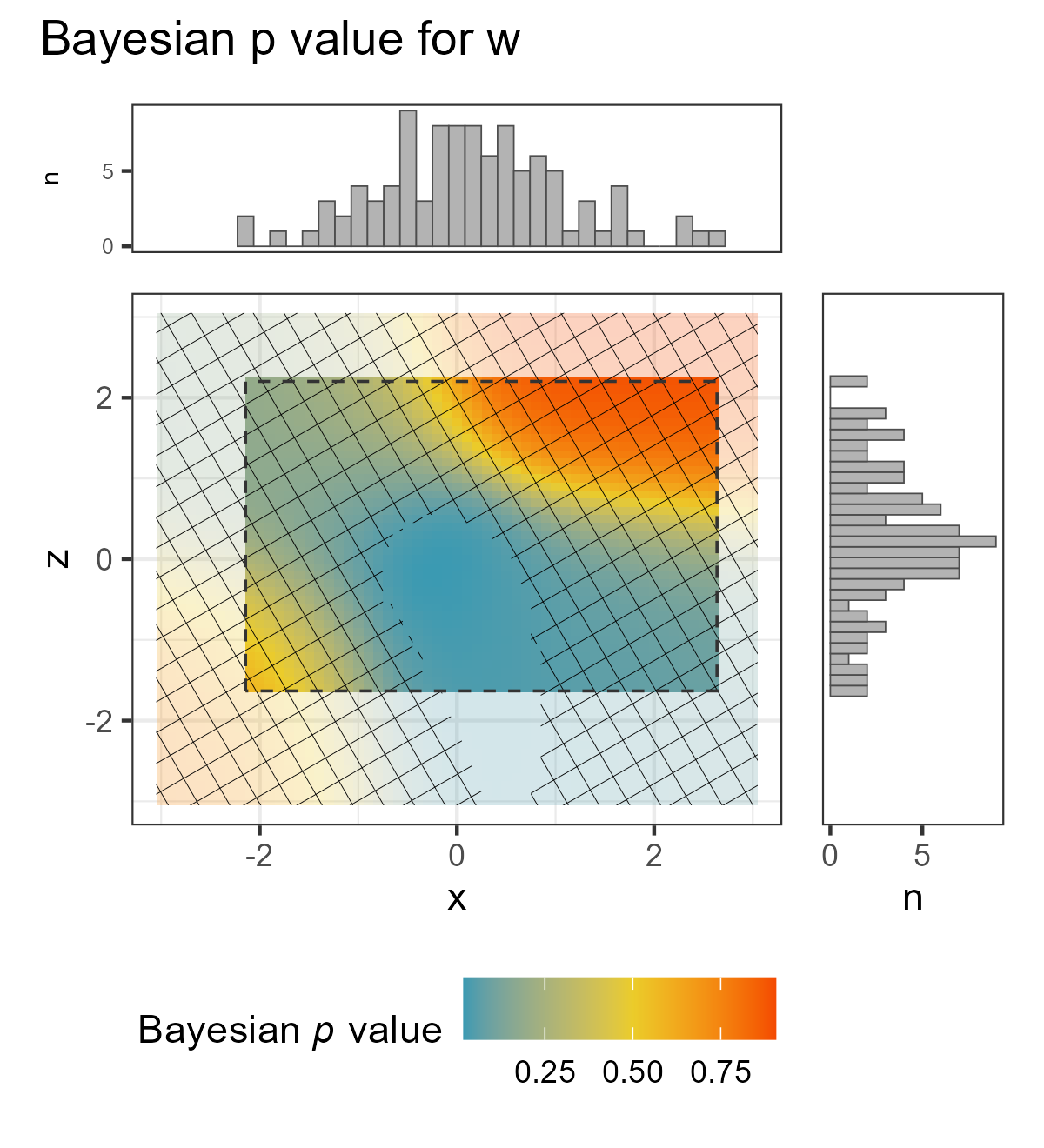}
  \caption{Bayesian $p$~value for the effect of \code{w} as a function of the
           moderators \code{x} and \code{z}.}
\end{subfigure}
\caption{JN3 heatmaps for the three-way interaction between \code{x},
         \code{z}, and \code{w}. Crosshatch overlays indicate cells where the
         Bayesian $p$~value falls inside the thresholds (here
         $0.05 < p < 0.95$).}
\label{fig:jn-bayes3}
\end{figure}

For three-way interactions the conditional posteriors are computed over the
grid $(m,w)$, producing heatmaps of both the posterior mean and the Bayesian
$p$~value, each crosshatched on cells falling inside the thresholds. Cells
whose Bayesian $p$~value lies in
$[\mathrm{threshold}_1, \mathrm{threshold}_2]$ are treated as inconclusive.
\begin{Schunk}
\begin{Sinput}
R> mod_bayes3 <- MCMCregress(y ~ x * z * w, data = dat,
+                            burnin = 1000, mcmc = 10000, thin = 1,
+                            verbose = 0)
R> jn_bayes3 <- JN(mod_bayes3, theta_1 = "x", theta_2 = "z", theta_3 = "w",
+                   theta_1_vals = seq(-3, 3, 0.1),
+                   theta_2_vals = seq(-3, 3, 0.1),
+                   theta_3_vals = seq(-3, 3, 0.1),
+                   thresholds = c(0.05, 0.95),
+                   support = list(dat$x, dat$z, dat$w))
\end{Sinput}
\end{Schunk}
Both heatmaps for the focal variable \code{w} are shown in
Figure~\ref{fig:jn-bayes3}; \code{type} selects between them.

\section{Stochastic actor-oriented models} \label{sec:saom}

Interactions in Stochastic Actor-Oriented Models (SAOMs), estimated by
\pkg{RSiena}'s \code{siena()} (formerly \code{siena07()}) \citep{RSiena} or by
\pkg{multiSiena} \citep{multiSiena}, between network statistics and actor
attributes are common yet difficult to interpret. Introducing SAOMs is beyond
the scope of this paper; see for instance \citet{Snijders:2017}. In short,
SAOMs are models for network evolution, designed to identify the structural and
covariate drivers behind the creation, dissolution, and maintenance of
connections in networks.

\code{JN()} handles \code{sienaFit} objects using integer indices referencing
the position of each effect in the \pkg{RSiena} effects object.\footnote{While
each effect in \pkg{RSiena} is named, within a given model effects are not
necessarily identifiable by their name alone. The same name can appear multiple times
for different dependent variables or effect types, making indexing by
integer easier than by name  to guarantee uniqueness.} For \code{sienaBayesFit} objects the
rate parameters are skipped in the indexing; if a position is out of range, the
error message lists the available effects with their positions. The moderator
ranges must be supplied and should reflect the empirical support of the
corresponding change statistics.

For Bayesian SAOMs, \code{JN()} automatically detects whether each required
parameter was estimated as a fixed effect ($\eta$, shared across groups) or as
a random effect ($\mu$, group-varying with a hyperparameter), and extracts the
appropriate draws; which of the two it used is recorded on the result and
reported by \code{print()}, \code{summary()} and the figures, so that the
choice made on the user's behalf is visible. With
\code{hyper_only = FALSE} and at least one randomly varying parameter involved,
it additionally loops over all $G$ groups and returns a \code{JN_list} of
group-specific analyses, which is useful for assessing heterogeneity of
interaction effects across networks. As for
\pkg{lme4} models, \code{jn_save()} is the practical way to inspect the
resulting figures.

\section{Graphical output and customization} \label{sec:graphics}

All figures are built from \pkg{ggplot2} objects \citep{Wickham:2016} and can
be modified with \pkg{ggplot2}'s layer system; where density panels are
attached they are composed with \pkg{patchwork} \citep{patchwork}. The
crosshatch pattern over non-significant regions is produced by \pkg{ggpattern}
\citep{ggpattern}. The colors for the high, mid, and low points of the
continuous scales are taken from the \code{Zissou1} palette of the
\pkg{wesanderson} package \citep{wesanderson}. Direct use of palettes is not
supported, but users can overlay their own by adding
\code{scale_fill_gradient2()} or \code{scale_color_gradient2()} post hoc (for
alternatives see \code{?ggplot2::scale_fill_gradient}).

Appearance settings are collected in \code{jn_style()}, which is passed to
\code{plot()} through its \code{style} argument, so that a house style can be
defined once and reused. Table~\ref{tab:colors} documents its arguments.

\begin{table}[t!]
\centering
\small
\begin{tabular}{p{3.5cm} p{7cm} l}
\toprule
Argument & Description & Default \\
\midrule
\code{sig_color}     & Ribbon fill for significant regions (two-way)
  & \cellcolor[HTML]{43CD80}\textcolor{black}{\code{"seagreen3"}} \\
\code{non_sig_color} & Ribbon fill for non-significant regions (two-way)
  & \cellcolor[HTML]{D2691E}\textcolor{white}{\code{"chocolate"}} \\
\code{line_color}    & Point-estimate line color (two-way)
  & \cellcolor[HTML]{000000}\textcolor{white}{\code{"black"}} \\
\code{color_low}     & Low end of diverging fill scale (heatmaps)
  & \cellcolor[HTML]{3B9AB2}\textcolor{black}{\code{"\#3B9AB2"}} \\
\code{color_mid}     & Midpoint color of diverging fill scale
  & \cellcolor[HTML]{EBCC2A}\textcolor{black}{\code{"\#EBCC2A"}} \\
\code{color_high}    & High end of diverging fill scale
  & \cellcolor[HTML]{F21A00}\textcolor{white}{\code{"\#F21A00"}} \\
\code{color_grid}    & Crosshatch line color
  & \cellcolor[HTML]{000000}\textcolor{white}{\code{"black"}} \\
\code{grid_density}  & Fraction of the crosshatched area covered by lines
  & \code{0.03} \\
\code{grid_spacing}  & Spacing between crosshatch lines
  & \code{0.05} \\
\code{grid_linewidth} & Outline weight of the crosshatch lines
  & \code{0} \\
\code{color_values}  & Within-cell text color
                       (heatmaps, $\le 7$ values)
  & \cellcolor[HTML]{666666}\textcolor{white}{\code{"grey40"}} \\
\code{crosshatch_non_sig}
                     & If \code{TRUE}, crosshatch marks the
                       non-significant region
  & \code{TRUE} \\
\code{show_density}  & Draw the data-density panels
  & \code{TRUE} \\
\code{density_fill}  & Fill of the density panels
  & \cellcolor[HTML]{B3B3B3}\textcolor{black}{\code{"grey70"}} \\
\code{density_color} & Outline of the density panels
  & \cellcolor[HTML]{4D4D4D}\textcolor{white}{\code{"grey30"}} \\
\code{density_height} & Size of the density panels relative to the main panel
  & \code{0.28} \\
\code{density_bins}  & Number of histogram bins
  & \code{30} \\
\code{extrap_alpha}  & Opacity of the margin beyond the observed moderator
                       values
  & \code{0.25} \\
\code{pos_color}     & Slopes and shading where the effect is significantly
                       positive (simple slopes)
  & \cellcolor[HTML]{F21A00}\textcolor{white}{\code{color\_high}} \\
\code{neg_color}     & Slopes and shading where the effect is significantly
                       negative (simple slopes)
  & \cellcolor[HTML]{3B9AB2}\textcolor{black}{\code{color\_low}} \\
\code{shade_alpha}   & Opacity of the shading between simple slopes
  & \code{0.35} \\
\code{slope_shading} & \code{"flat"} or \code{"density"}: shade by the
                       moderator's distribution
  & \code{"flat"} \\
\code{mode_color}    & Slope at the most common moderator value
  & \cellcolor[HTML]{4D4D4D}\textcolor{white}{\code{"grey30"}} \\
\bottomrule
\end{tabular}
\caption{\label{tab:colors}Arguments of \code{jn\_style()}, which controls the
appearance of every figure in the package. Default color values are
highlighted in their respective color.}
\end{table}


\section{Summary and outlook} \label{sec:conc}

\pkg{int3ract} provides a unified interface for the Johnson--Neyman technique
and its three-way extension across a broad class of frequentist and Bayesian
models. A single generic, \code{JN()}, dispatches on the fitted object and
returns a classed result whose \code{summary()} reports the regions of
significance and whose \code{plot()} draws them. Because all model-specific
knowledge is confined to \code{jn_input()} methods, extending the package to a
new estimator is a matter of a few lines rather than a change to its internals.

Researchers are encouraged to report not only the JN boundaries but also the
full plots as supplementary material, in keeping with calls for more
transparent and complete reporting of interaction effects
\citep{Simonsohn+Simmons+Nelson:2020}.

\pkg{int3ract} is available from CRAN. Readers who want a shorter route into
the package than this manuscript can start from its vignette,
\code{vignette("int3ract")}, which covers the same ground as
Sections~\ref{sec:freq} to~\ref{sec:graphics} in condensed form---from a
two-way analysis with \code{lm()} through the three-way moderator grid,
Bayesian input, graphical customization and the \code{jn_input()} extension
point.


\section*{Computational details}

The results in this paper were obtained using \proglang{R}~4.6.1 with
\pkg{int3ract}~2.2.0, \pkg{ggplot2}~4.0.3, \pkg{ggpattern}~1.3.1,
\pkg{patchwork}~1.3.2 and \pkg{MCMCpack}~1.7.1. The \pkg{lavaan} examples of
Section~\ref{sec:freq} were checked against \pkg{lavaan}~0.7.2.
\proglang{R} itself and all
packages used other than \pkg{RSiena} and \pkg{multiSiena} are available from
the Comprehensive \proglang{R} Archive Network (CRAN) at
\url{https://CRAN.R-project.org/}; \pkg{RSiena} and \pkg{multiSiena} are
available from \url{https://www.stats.ox.ac.uk/~snijders/siena/}. All
computations are deterministic given the seed set at the start of the
replication script, apart from the Markov chain Monte Carlo runs in
Section~\ref{sec:bayes}, which are seeded by the same call.


\bibliography{refs}

\end{document}